\documentclass[a4paper,fleqn,usenatbib]{mnras}
\usepackage{newtxtext,newtxmath}
\usepackage[T1]{fontenc}
\usepackage{ae,aecompl}
\usepackage{siunitx}
\usepackage{multicol}
\usepackage{multirow}
\usepackage{graphicx}	% Including figure files
\usepackage{amsmath}	% Advanced maths commands
\usepackage{amssymb}	% Extra maths symbols
\usepackage{subfloat}
\usepackage{lipsum}
\usepackage{caption}
\usepackage{ctable}
\usepackage{enumitem}
\usepackage{latexsym, amscd, psfrag}
\usepackage{epsfig,color}
\usepackage{epstopdf,subfig}

\title[UV and optical view of galaxies in the Coma SCL]{Ultraviolet and optical view of galaxies in the Coma Supercluster}

\author[Mahajan et al.]
{Smriti Mahajan\thanks{E-mail: smritimahajan@iisermohali.ac.in}, 
Ankit Singh, Devika Shobhana 
\vspace{0.4cm}\\
Indian Institute for Science Education and Research Mohali - IISERM, Knowledge City, Manauli, 140306, Punjab, India }

\def\g{{\it GALEX}}
\def\nuv{{\it NUV}}
\def\fuv{{\it FUV}}
\def\smass{{$M^*$}}
\def\nuvr{{$NUV-r$}}
\def\fnuv{{$FUV-NUV$}}
\def\disperse{{\sc DisPerSe}}

\definecolor{grey}{rgb}{0.5,0.6,0.7}

\definecolor{amber}{rgb}{1.0,0.49,0.0}

\date{Accepted XXX. Received YYY; in original form ZZZ}

\pubyear{}

\begin{document}
\pagerange{\pageref{firstpage}--\pageref{lastpage}}
\maketitle

\label{firstpage}

% Abstract of the paper
\begin{abstract}
 The Coma supercluster ($100 h^{-1}$Mpc) offers an unprecedented contiguous range of environments in the nearby Universe. In this paper 
 we present a catalogue of spectroscopically confirmed galaxies in the Coma supercluster detected in the ultraviolet (UV) 
 wavebands. We use the arsenal of UV and optical data for galaxies in the Coma supercluster covering $\sim 500$ square degrees on the 
 sky to study their photometric and spectroscopic properties as a
 function of environment at various scales. We identify the different components of the cosmic-web: large-scale filaments
 and voids using Discrete Persistent Structures Extractor, and groups and clusters using Hierarchical Density-based spatial clustering of 
 applications with noise, respectively. We find that in the Coma supercluster the median emission in H$\alpha$ inclines, 
 while the $g-r$ and \fnuv~colours of galaxies become bluer moving further away from the spine of the filaments out to a radius of $\sim 1$ Mpc. On the 
 other hand, an opposite trend is observed as the distance between the galaxy and centre of the nearest cluster or group decreases. 
 Our analysis supports the hypothesis that properties of galaxies are not just defined by its stellar mass and large-scale density, but also by the 
 environmental processes resulting due to 
 %tidal interactions with 
 the intrafilament medium whose role in accelerating galaxy transformations needs to be investigated thoroughly using multi-wavelength data.  
\end{abstract}


%%%%%%%%%%%%%%%%%%%%%%%%%%%%%%%%%%%%%%%%%%%%%%%%%%
%%%%%%%%%%%%%%%%% BODY OF PAPER %%%%%%%%%%%%%%%%%%

\section{Introduction}

 Large-scale cosmic-web filaments are crucial, yet poorly investigated intermediate density environments capable of accelerating galaxy
 transformation \citep{porter08,haines11,alpaslan16,martinez16,kuutma17}. Studies at high redshift \citep{fadda08,geach11,coppin12,darvish14}
 as well as $z\sim0$ \citep[e.g.][]{porter08,mahajan10,haines11,alpaslan16,kim16,kleiner17,kuutma17,kraljic18} have now shown that 
  the intermediate density environment prevalent in filaments not just bridge the gap between
 the dense interior of clusters and `voids' devoid of galaxies, but play a rather important role in modulating galaxy observables. Owing to the 
 highly coherent flow of galaxies along the filaments \citep{gonzalez09}, they have not just been observed at all wavelengths 
 from ultraviolet (UV) \citep{gallazzi09,geach11,coppin12} to infrared \citep{fadda08,haines08} by direct methods such as quantifying the 
 distribution of galaxies, thermal Sunyaev-Ze{'l}dovich (tSZ) effect \citep{bonjean18} and weak gravitational 
 lensing \citep{dietrich12,jauzac12}, but also indirectly using a bent double lobe radio source \citep{edwards10b}.

 Recently, filaments, in particular the warm hot intergalactic medium (WHIM) in filaments has been the pivot of several studies. Using 
 emission in the soft x-ray bands, such studies estimate the WHIM temperature in filaments to be $\sim$3--8 keV 
 \citep{eckert15,akamatsu17,parekh17,tanimura17}. Furthermore, the cosmic microwave background map from the Planck together with
 the Canada France Hawaii Telescope Lensing Survey, as well as the Two-Micron All-Sky Redshift Survey of galaxies 
 suggest that at least half of the missing baryons in the Universe may reside as WHIM in large-scale filaments tracing the
 dark matter distribution \citep{van14,genova15}. Therefore, undeniably it is crucial to characterise the large-scale structure (LSS) of the
 Universe and comprehend the impact of the cosmic-web on properties of galaxies. 
  
 The Coma supercluster is one of the richest large-scale structure \citep{chincarini76} in the nearby Universe comprising two clusters
 of galaxies, connected by a web of large-scale filaments around $30 h_{70}^{-1}$ Mpc long \citep[e.g.][]{fontanelli84}. The two clusters, 
 Coma (Abell 1656) and Abell 1367, along with the filaments of galaxies, dispersed with several small galaxy groups span $\sim 500$ square
 degrees on the sky \citep{mahajan10}. The large-scale filaments in the Coma supercluster have not just been observed by means of the  
 galaxy distribution in the optical wavebands \citep{gregory78,mahajan10}, but also diffuse emission in the radio continuum \citep{kim89}.
 
 Studies of clusters and groups at $z\sim0$ have evidently shown that outskirts of groups and clusters 
 \citep{zabludoff98,rines05,wang04,cortese07,tran09,gavazzi10,smith10,sun10,coppin11,mahajan12,verdugo12,mahajan13} and filaments 
 of galaxies \citep{porter07,boue08,fadda08,porter08,edwards10,biviano11} are favourable sites for galaxy transformations. 
 Based on a study using optical data from the Sloan Digital Sky Survey (SDSS) data release (DR)~7, \citet{mahajan10} found that the star 
 formation-density relation in the Coma supercluster for the giant galaxies is much weaker than their dwarf counterparts. However,
 the fraction of star-forming galaxies for both declines to $\sim 0$ at the centre of the clusters \citep[also see][]{mahajan11}. 
 \citet{cybulski14} furthered the study of star formation in the Coma supercluster
 by combining a complementary optical dataset from SDSS DR~9, with IR data from the {\it Wide-Field Infrared Survey Explorer}
  \citep[WISE;][]{wright10} and UV data from the {\it Galaxy Evolution Explorer} \citep[\g;][]{martin05}. 
 \citet{cybulski14} corroborated the results of \citet{mahajan10,mahajan11} by probing both obscured and unobscured star formation
 down to $\sim 0.02$ M$_\odot$ yr$^{-1}$, in order to quantify the effect of different types of large-scale environments: groups, clusters, filaments 
 and voids, on quenching star formation (SF) in galaxies.   
   
 In the absence of dust in star-forming galaxies, the UV emission is a good tracer of massive ($>10 M_{\odot}$) star formation. On the 
 other hand, optical emission lines such as H$\alpha$ probe instantaneous star formation over a time-scale of $\lesssim 20$ Myr \citep{kennicutt98}.
 Assuming that the UV luminosity is not overwhelmed by contribution from the old stellar populations due to the UV upturn 
 such as in massive early-type galaxies \citep{oconnell99}, the UV luminosity measures star formation over a timescale of $\sim 100$ 
 Myr \citep{kennicutt98}. Hence, the star formation rate (SFR) estimated from optical emission lines delineates the continuous SF in a 
 galaxy, while the SFR determined from the UV is representative of its recent SF activity. 
 But even though \g~and its predecessor UV imagers have been used to investigate individual galaxies within clusters and groups 
 \citep[e.g.][]{hicks05}, or galaxy populations therein \citep[e.g.][]{donas90,donas95,cornett98,boselli05b}, limited work has been done
 to analyse the UV properties of galaxies in the large-scale cosmic-web.
 
 Since the Coma supercluster is one of the most well studied regions in the nearby Universe, many other authors 
 \citep[e.g.][]{bernstein95,mobasher03,hammer12,smith10,smith12} have made use of optical and UV data to study the 
 Coma and Abell~1367 clusters and their surroundings. With the advent of large redshift surveys several studies   
 \citep{gavazzi10,mahajan10,mahajan11,cybulski14,gavazzi13} have also used multi-wavelength
 data at optical, UV and 21 cm continuum to study the properties of galaxies in the entire supercluster region. In this paper we make use 
 of similar datasets: UV data derived from \g~and optical spectroscopic and photometric data from the SDSS for
 the entire Coma supercluster to further explore the impact of environment on the properties of galaxies.
 
 Conventionally, the `environment' of galaxies is quantified as the projected density of galaxies in a fixed 2-d or 3-d region of the sky 
 \citep[e.g.][]{dressler80}. \citet{muldrew12} combined 20 published methods of defining environment into two (i) methods which use
 nearest-neighbours to probe the underlying galaxy density and, (ii) fixed aperture methods. \citet{muldrew12} found that while the former are
 better suited for quantifying internal density within massive halos, the latter, fixed-aperture methods are better for probing the large-scale
 environment. Therefore, in order to characterise the large-scale cosmic-web, a combination of these methodologies is required to quantify 
 the environment on different scales. In this work we implement their result by making use of two different algorithms to define the large-scale
 filaments, and high density nodes of the cosmic-web characterised as clusters and groups.
 We also present a catalogue of all spectroscopically-confirmed galaxy members of the Coma supercluster detected in the UV. 
 
 This paper is organised as follows: in the next section we describe our datasets,
 followed by definition of environment in Sec.~\ref{s:env}. In Sec.~\ref{colour} we analyse the broadband colours of galaxies as a function
 of environment, while in Sec.~\ref{web} we study the impact of the large-scale filaments on the properties of galaxies. Finally, we discuss our
 results in the context of the existing literature in Sec.~\ref{discuss} and summarise our results in Sec.~\ref{results}. Throughout this paper
 we use concordance $\Lambda$CDM cosmological model with $H_0=70$\,km s$^{-1}$ Mpc$^{-1}$, $\Omega_\Lambda=0.7$ and 
 $\Omega_m=0.3$ to calculate distances and magnitudes. We note that at the redshift of the Coma cluster ($z=0.023$) our results are 
 independent of the cosmological model used.    
 
\section{Data}

 This work is based on the optical photometric and spectroscopic data acquired from the SDSS data release 12 \citep{alam15}, and UV data
 from the \g~survey \citep{bianchi14}. In the following we describe both the datasets used in this work.
 
 \subsection{Optical data}
 
 The Sloan Digital Sky Survey (SDSS) uses two fibre-fed double spectrographs, covering a wavelength range of 
 3800--9000~\AA. The resolution $\lambda/\Delta \lambda$ varies between 1500 and 2500 in different bands. Galaxy 
 magnitudes\footnote{Throughout this paper we use the SDSS `model' magnitudes. The model magnitude for each galaxy 
 is calculated using the best-fit parameters in the $r$ band, and applied to all other bands; the light is therefore measured
  consistently through the same aperture in all bands, allowing for an unbiased measure of galaxy's colours.} are $K$-corrected to $z=0$  
 \citep{chilingarian12}, and for galactic extinction. For the latter, the $E(B-V)$ values in each of the $g$ and $r$ bands are calculated using 
 the Schlegel (1998) dust maps. 
   
 Following \citet{mahajan10}, we select all galaxies brighter than $m_r=17.77$ mag and spectroscopic data which lie within  
 $170^\circ \leq \rm{RA} \leq 200^\circ, 17^\circ \leq \rm{Dec} \leq 33^\circ$ and $0.013 \leq z \leq 0.033$. These criteria yield
 a total of 4,280 galaxies. In this work we use the equivalent width of the H$\alpha$ emission line as an indicator of the instantaneous 
 ($< 10$ Myr) star formation rate (SFR) of galaxies. 
 
 The spectra of 86\% of all the galaxies in the Coma supercluster have all four emission lines (H$\alpha$, [NII], H$\beta$ and [OIII]) 
 required to confirm the presence of nuclear activity based on the emission-line ratio diagnostics as first proposed by \citet{bpt}. For this work, 
 we use the classification criteria of \citet{kewley01} to identify 518 ($\sim 12\%$) galaxies as AGN. Where applicable, we perform the analysis using
 non-AGN galaxies only i.e. the entire sample minus the 518 galaxies whose spectra is dominated by AGN emission. 

\subsection{Ultraviolet data}

 \begin{figure*}
 \centering{
 {\rotatebox{270}{\epsfig{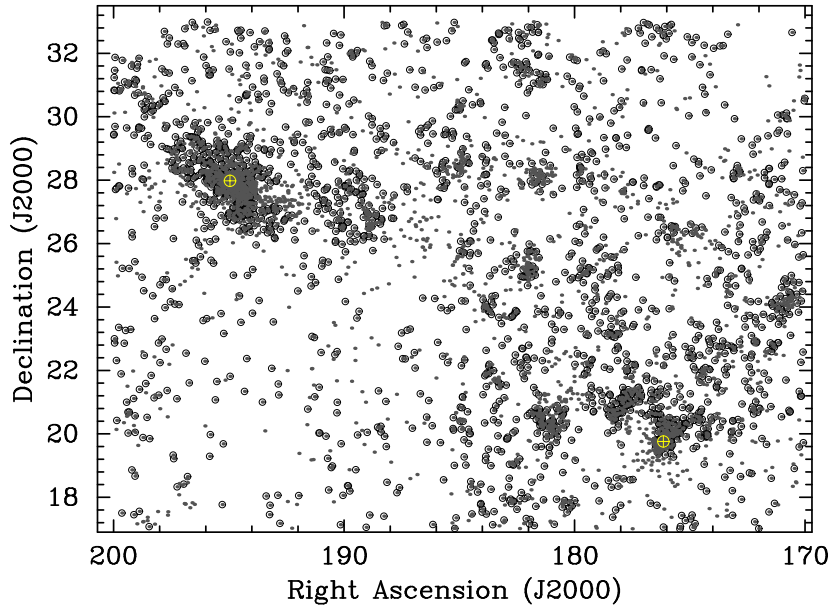}}}}
 \caption{The Coma supercluster. All spectroscopically confirmed galaxies are represented as {\it dots}, and all
 galaxies with a UV counterparts are {\it encircled}. The {\it yellow encircled crosses} represent the centre of the Coma and Abell 1367 
 clusters.}
 \label{scl}
 \end{figure*}

\begin{table*}
\caption{The SDSS and \g~data for all the Coma supercluster galaxies in our sample. The local environment of galaxies listed
 in column 8 is:  0: void, 1: filaments and 2: cluster/group, respectively. See section \ref{s:env} for detailed definition of these environments.}
\begin{tabular}{|c|c|c|c|c|c|c|c|c|c|c|c|c|}
\hline
 Plate & MJD & Fiber & \g~ObjID & RA & Dec & z & Local & $M_r$ & $m_{NUV}$ & $\Delta m_{NUV}$ & $m_{FUV}$ & $\Delta m_{FUV}$ \\
          &          &              &           & (J2000)  & (J2000)   &    &   env   & mag    & mag   & mag   & mag    & mag  \\
\hline
2241 & 54169 & 562        & - &       195.547 &  28.172 & 0.0299 & 2 & -18.69 & - & - & - & -\\
2241 & 54169 & 517 &  3191682282023817108 & 194.984 &  27.746 & 0.0298 & 2 & -19.43 &  19.05 &   0.02 &  19.69 &   0.03\\
2240 & 53823 & 530        & - &       194.629 &  28.234 & 0.0243 & 2 & -19.18 & - & - & - & -\\
2240 & 53823 & 617 &  3187917554210322283 & 195.170 &  27.997 & 0.0236 & 2 & -20.67 &  20.61 &   0.02 &  22.07 &   0.06\\
2509 & 54180 & 226        & - &       176.020 &  19.533 & 0.0243 & 2 & -18.63 & - & - & - & -\\
2506 & 54179 & 548 &  2555795125061031411 & 176.134 &  20.107 & 0.0240 & 2 & -20.05 &  19.87 &   0.04 &  20.67 &   0.07\\
2242 & 54153 & 384 &  3054146571327246615 & 195.313 &  27.669 & 0.0236 & 2 & -18.64 &  22.01 &   0.15 &  23.13 &   0.38\\
2241 & 54169 & 573        & - &       195.406 &  28.016 & 0.0238 & 2 & -18.07 & - & - & - & -\\
2241 & 54169 & 371        & - &       194.523 &  28.243 & 0.0240 & 2 & -21.39 & - & - & - & -\\
2240 & 53823 & 518        & - &       194.548 &  27.940 & 0.0284 & 2 & -19.45 & - & - & - & -\\
\hline
\end{tabular}
\label{opt-uv-cat}
\end{table*}

 The ultraviolet photometric data for this work is taken from the {\it{Galaxy Evolution EXplorer}} final data release 
 \citep[{\it{GALEX}};][]{bianchi14}. {\it{GALEX}} conducted an all-sky imaging survey along with targeted programs in two
 photometric bands: 1516\,\AA\ (``far ultraviolet'' or FUV) and 2267\,\AA\ (``near ultraviolet'' or NUV). The bulk of our 
 sample consists of bright, nearby galaxies, and therefore no exposure time or brightness limit constraints were imposed while looking for a
 {\it{GALEX}} detection. For best photometric and astrometric quality we have restricted our sample to the inner $0.5^\circ$ of the \g~field of view and having NUV\_artifact $< 1$. As suggested by \citet{wyder07}, the latter excludes objects whose flux may be contaminated by reflection from bright stars.
 
 65\% of the UV imaging data used in this paper were taken as part of the
 {\it{GALEX}}'s primary All-Sky Imaging Survey (AIS) with an effective exposure time of ${\sim}0.1$\,ks. Most of the rest of the data comes
 from individual guest investigator programs (28\%) or the Nearby Galaxies Survey (NGS; 6\%) with an effective exposure time of
 ${\sim}1.5$\,ks. The remaining data was taken as part of {\it{GALEX}}'s medium or deep imaging surveys (MIS and DIS), having an 
 average exposure time of ${\sim}1.5$ and $30$\,ks respectively. 

 We searched for the UV counterpart of each spectroscopically-confirmed supercluster galaxy within a circular aperture with radius
 4$^{\prime\prime}$ centred on the optical source \citep{budavari09}. Amongst others, this process also yields 477 optical sources multiply matched
 to many UV sources. Following \citet{bianchi11}, we considered the multiply-matched UV sources as duplicates if they were within
 $2.5^{\prime\prime}$ of each other. Of those UV sources which satisfied this criteria, if the {\it photoextractid} of both the UV 
 sources is the same implying both of them are from the same observation, they were both considered unique, else they were
 assumed to be multiple observations of the same source. In the former case, the matched counterpart was represented by the
 UV source with the highest \nuv~exposure time. In case of equal \nuv~exposure times for both, the one closest to the field centre in the
 {\it{GALEX}} image was chosen. 
 
For the \fuv~and \nuv~photometry we used Kron magnitudes generated by the \g~pipeline and corrected for galactic extinction using Schlegel (1998) dust maps, assuming the  reddening law of \citet{cardelli89}. The reddening factor for \g~bands is $AFUV/E(B-V) = 8.24$ mag and $ANUV/E(B-V) = 8.2$ mag  \citep{wyder07}, respectively, which lead to a median extinction correction of $\sim 0.23$ mag in both the UV bands. 
 Since \g~colour is an important part of our analysis, we chose the \nuv~band as the primary band and then obtained the corresponding \fuv~magnitude measured within the same Kron aperture as estimated using the \nuv~image to get an accurate UV colour. 

 \citet{hammer10} have created a source catalogue of objects detected in a 26 ks deep \g~field in the Coma cluster. For 94 galaxies (matched within $10^{\prime\prime}$) of their sources, we find a median offset of 0.1 mag in the \fuv~and 0.2 mag in the \nuv~magnitudes, respectively. 
 The discrepancy in the measured magnitudes is due to the bayesian deblending
 technique used by these authors, since the photometry thus obtained is relatively more sensitive to the apparent sizes of galaxies, and  
 underestimates the flux for extended ($R_{90,r}>10^{\prime\prime}$) galaxies \citep{hammer10}, which make up $> 50\%$ of our sample.
 
 Our final UV catalogue comprises 2,447 UV sources uniquely matched to spectroscopically-confirmed SDSS galaxies in the
 Coma supercluster region. The combined optical-UV data are shown in Figure~\ref{scl}, and the data for all 4,280 galaxies are 
 given in Table~\ref{opt-uv-cat}. The columns in Table~\ref{opt-uv-cat} are: (i) Plate, (ii) MJD, (iii) Fibre ID, (iv) \g~ID, (v) Right ascension (J2000), 
 (vi) Declination (J2000), (vii) redshift (viii) local environment (0: void, 1: filaments and 2: cluster/group), (ix) $r-$band magnitude (x) $FUV$ magnitude, 
 (xi) error in $FUV$ magnitude, (xii) $NUV$ magnitude and (xiii) error in $NUV$ magnitude. The first three columns can be used to crossmatch 
 galaxy data with the SDSS database, while the unique \g~ID can be used for the same with the \g~database. 
 
\begin{figure}
 \centering{
 {\rotatebox{270}{\epsfig{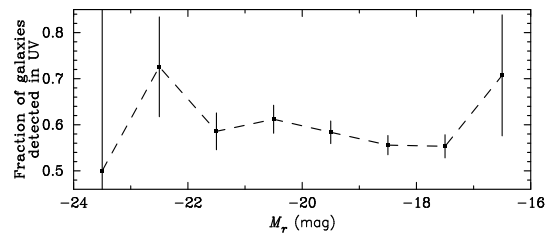}}}}
 \caption{Fraction of Coma supercluster galaxies with a UV counterparts in bins of absolute $r$-band magnitude. On average,
 $\sim 60\%$ of the galaxies are detected by \g~in each bin of $M_r$. The error bars on each point represent Poissonian uncertainty. }
 \label{uv-frac}
 \end{figure}

 Since we are covering such a large chunk of the sky ($\sim 500$ sq. degrees), it is unsurprising that
 the depth of \g~observations varies widely over the entire supercluster. \citet[][see their sec.~2.2 for details]{cybulski14}, have shown that the
 \g~data for the Coma supercluster is 75\% complete to a luminosity of $\sim 26.42$ ergs. We therefore stress that our UV catalogue
 is incomplete for two major reasons: firstly, due to incomplete spatial coverage of \g~due to the presence of bright foreground sources
 (e.g. region around $\alpha\sim186^\circ$ and $\delta\sim26^\circ$ in Figure~\ref{scl}). 
 Secondly, as shown in Figure~\ref{uv-frac} only $\sim 60\%$ of the spectroscopically-confirmed galaxies are detected in each
 bin of $r-$band magnitude. However, since the fraction of galaxies missing a UV counterpart is almost uniform across the entire
 luminosity range, except the lowest and the highest luminosity bin comprising $2$ and $41$ galaxies respectively, we have not
 corrected for this incompleteness in the following analysis. We use all the UV-detected galaxies for all UV-based analysis, and all 
 spectroscopically-confirmed galaxies for all the optical analysis below.
 
 \section{The environment}
 \label{s:env}
 
 It is now well established that the Coma supercluster comprises of a filamentary network including the two major clusters Coma and Abell~1367,
 and other smaller galaxy groups \citep{gregory78,mahajan10}.
 This filamentary network has been mapped by the distribution of galaxies as well as observed in the radio continuum at 326 MHz \citep{kim89}. 
 In the following sections we will analyse optical and ultraviolet properties of galaxies in the Coma supercluster as a function of their environment. 
 In order to do so, we first modelled the complex cosmic-web of the Coma supercluster by identifying its various components as described below. 
  
 \subsection{The cosmic-web with \disperse}
 \label{disperse}
 %%%%%%%%% DISPERSE DETAILS BY ANKIT %%%%%%%%%%%%%%%%%%
%%%%%%%%%%%%%%%%%%%%%%%%%%%%%%%%%%%%%%%%%%%%%

   \begin{figure*}
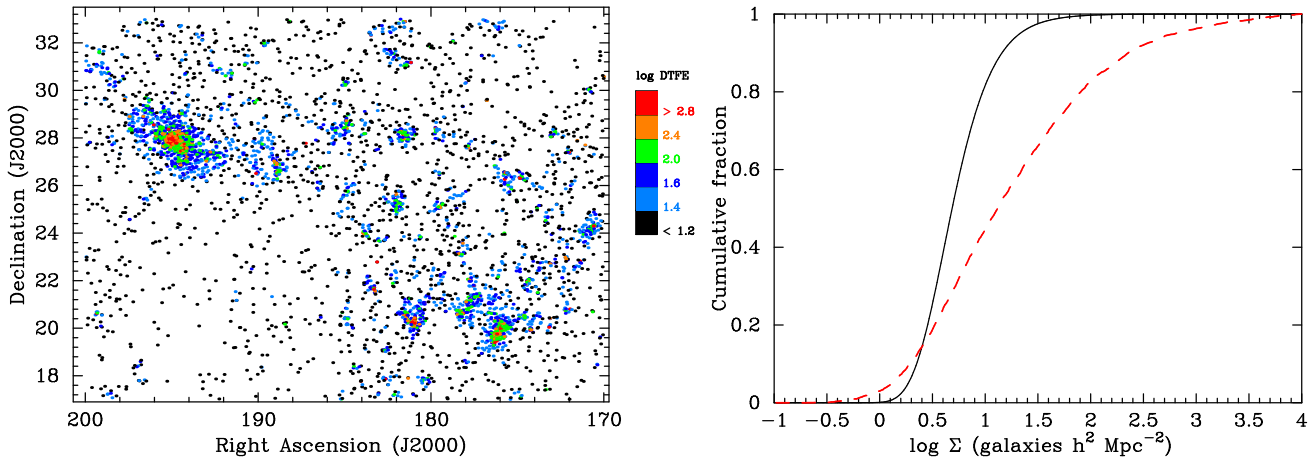

 \centering{
 {\rotatebox{270}{\epsfig{file=sky-dtfe.ps,width=6cm}}}}
 \vspace{5mm}
 \centering{
 {\rotatebox{270}{\epsfig{file=cumulative-fraction.ps,width=6cm}}}}
 \caption{{\it (left:)} The Coma supercluster galaxies colour-coded with log of Delaunay Tessellation Field Estimator value (see text). {\it (right:)}
 The mean cumulative distribution of the DTFE density for the random samples {\it (black solid line)} compared with the same for the Coma 
 supercluster {\it (red dashed line)}. The Coma supercluster region is not just denser relative to the random distributions, but also comprise very 
 over-dense (clusters and groups) and under-dense (voids) regions. }
 \label{disp}
 \end{figure*}

 The filaments in the Coma supercluster in this work are defined using Discrete Persistent Structures Extractor \citep[\disperse;][]{sousbie11},
 which is an algorithm capable of finding structures, in particular linear structures such as large-scale filaments within discrete 2d or 3d datasets
 \citep[see][for illustrations of \disperse~in the astrophysical context]{sousbie11b}. \disperse~allows us to work directly with particles, and
 requires only one tunable parameter corresponding to the significance of the retained features in units of $\sigma$\footnote{In this work
 we use a significance threshold of $3\sigma$ which corresponds to a probability of $\sim 0.997$ \citep{sousbie11} .}. 
 Furthermore, \disperse~does not make any presumptions about the quality of sampling, homogeneity or topology of the space.
 
 \disperse~is based on discrete Morse theory and persistence homology theory. The Delaunay tessellation is used to segregate the whole space
 into triangular regions. The Delaunay Tessellation Field Estimator \citep[DTFE;][]{schaap00,cautun11} allows the estimation of density
 at each vertex of the triangular complexes created by Delaunay tessellation. The estimated density can then be used in the Morse theory 
 to identify the critical points where the gradient of density field vanishes.

 The topology of a function can be studied using the topology of points which are above a continuous monotonically changing threshold. 
 If the persistence value of a pair of critical points is higher than the noise level defined by the root mean square of persistence values of 
 randomly selected points, the detected feature tends to have a significant density contrast. One can select a threshold of persistence values 
 above which the pairs are to be considered for analysis. This enables extraction of all significant features in the space while eliminating noise. 
 The filaments can be identified as lines connecting two maximas residing at the centres of groups and cluster regions. 
  For a detailed discussion of the theoretical framework which forms the basis for \disperse, we refer the reader to \citet{sousbie11}.
    
 In Figure~\ref{disp} (left) we show the position of all galaxies colour-coded by their DTFE value. Following \citet{cybulski14},
 we generated 1500 random samples of galaxy densities with the same number of galaxies as in the Coma supercluster, but their 
 positions distributed randomly. Figure~\ref{disp} (right) shows a comparison between the cumulative distribution of the
 mean of the galaxy density of 1500 random samples with the Coma supercluster. This figure is directly comparable to figure~4 (right)
 of \citet{cybulski14}. Since we have used a dataset very similar to theirs, we also observe trends similar to those seen by \citet{cybulski14}.
 The densities in the Coma supercluster gradually increases from a few to $\sim100-1000$ galaxies per square Mpc. 
 
 Figure~\ref{disp} shows that
 the Coma supercluster on average is denser than a randomly selected region of the Universe, in the sense that the median density of 
  galaxies in the supercluster is $\sim 0.4$ dex higher relative to the mean of random samples. These observations are well in agreement with
 other studies \citep{gavazzi10}, reporting that with a mean density of $0.019$ gal $(h^{-1} Mpc)^{-3}$ the Coma supercluster is three times over-dense 
 relative to the Universe in general \citep[0.006 gal $h^{-3} Mpc^{-3}$ for $M_i<-19.5$ mag;][]{hogg04}.

  \subsection{Identifying Galaxy Groups}	
 \label{groups}
 
 Along with the clusters and the cosmic-web, another prominent environment in the Coma supercluster is of galaxy groups. In this work we identify 
 the galaxy groups by Hierarchical Density-Based Spatial Clustering of Applications with Noise \citep[{\sc hdbscan}\footnote{Specifically we 
 adopted the Python implementation of {\sc hdbscan}};][]{mcinnes17b}. 
 {\sc hdbscan} is a theoretically and practically improved version of the popular Density-based spatial clustering of applications with noise 
 ({\sc dbscan}) algorithm \citep{ester96,schubert17}.
 
 {\sc dbscan} requires minimal domain knowledge to determine input parameters, and works efficiently on large datasets. 
 For our work, {\sc dbscan} is a better choice relative to other unsupervised clustering algorithms (e.g. k-means) because of its ability to
 detect arbitrarily shaped clusters. The functioning of the algorithm is primarily governed by  
 two parameters, minPts (minimum number of data points to be considered as a `cluster') and $\epsilon$ (the maximum radius permissible for a cluster).
 To begin with, {\sc dbscan} classifies the data into three categories: core points (points with the maximum number of minPts in their vicinity),
 border points (points with lesser number of minPts but in the vicinity of a core point), and outliers (all other data points). The main issue
  with this approach is %{\sc dbscan} is that it is not deterministic, 
 that the points which are in the vicinity of more than one core points, will have equal probability of being assigned to either of the clusters.
% any of the cluster depending on which of them is processed first. 
 Another drawback of the {\sc dbscan} algorithm is that it is inefficient in finding clusters in an inhomogeneous dataset, making it very difficult to select
 the initial parameters for the algorithm.

 {\sc hdbscan} improves upon the methodology of the {\sc dbscan} algorithm by extracting flat clustering based on stability of clusters \citep{campello13}.
 The main governing clustering parameter in {\sc hdbscan} is min\_cluster\_size (the minimum size of an allowed cluster), which is chosen by the 
 user based on the knowledge of the dataset. Another parameter which is crucial in deciding the final clustering is min\_samples, which
 sets how conservative the clustering should be. By default min\_samples is set equal to min\_cluster\_size, but can be tuned to suit the dataset. 
 A very large value of min\_samples will label most of the data points as noise, where as setting it as unity will allow
 all the data points to be clustered. In this work we adopt min\_sample\_size $= 20$ and min\_samples $= 10$.
 The working of {\sc hdbscan} can intuitively be divided into five steps:
 \begin{itemize}
 \item calculation of the {\it mutual reachability distance} \citep{campello15} for all the data points, where, the mutual reachability distance between
 two points $a$ and $b$ is defined as the maximum of the distance to the $k$th nearest neighbour for $a$, $b$ or the distance between 
 $a$ and $b$, i.e. $d_{mutual} (a,b) = max\{{\rm core}_k(a),{\rm core}_k(b), d(a,b)\}$,
 \item construction of the minimum spanning tree of the data points weighted by their mutual reachability distance,
 \item building the hierarchy of the connected components,
 \item depending on the minimum size of the cluster allowed (min\_cluster\_size), {\sc hdbscan} condenses and retains the 
  clusters having size greater than min\_cluster\_size. This step significantly reduces the number of branches relative to the previous step; and
 \item the clusters which persist longer than their descendant clusters in the hierarchy tree are chosen as stable clusters and their descendants are
 disregarded, else the descendant clusters are chosen as stable clusters. 
 \end{itemize} 
 The final step allows {\sc hdbscan} to select clusters with varying density. 

  \begin{figure*}
 \centering{
 {\rotatebox{270}{\epsfig{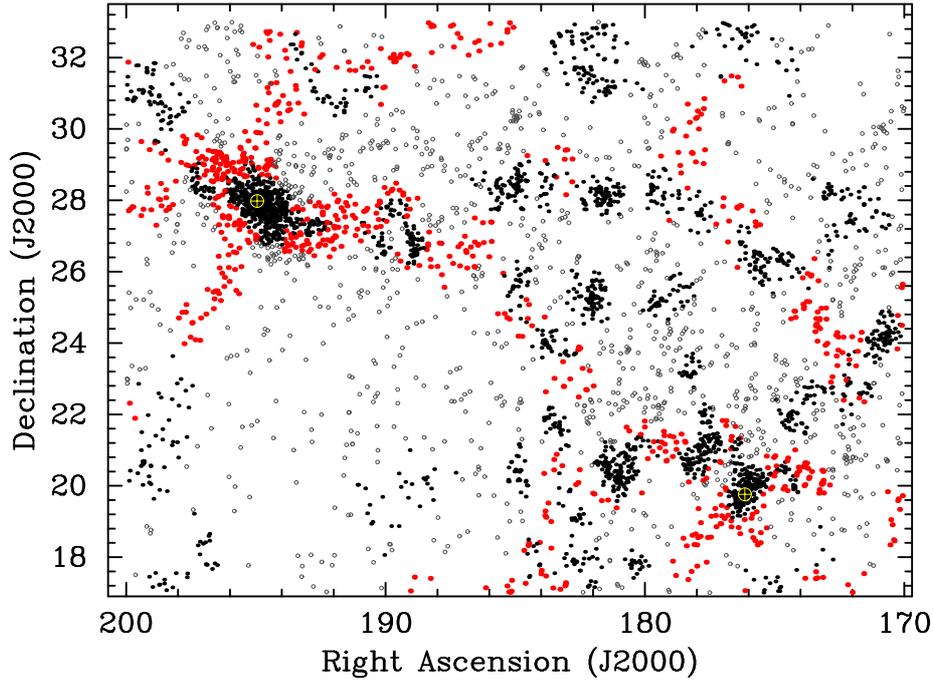}}}}
 \caption{Different environments in the Coma supercluster. The {\it grey open circles, red points}, and {\it black points} represent the void, 
 cosmic-web and cluster galaxies, respectively. The {\it yellow encircled crosses} represent the centre of the Coma and Abell 1367 clusters. }
 \label{env}
 \end{figure*}

 \subsection{Characterisation of environment}
 \label{ss:env}
 
 For further analysis of galaxy properties as a function of their environment, we segregate the Coma supercluster into three `local' environments:
 \begin{itemize}
 \item Cluster/group galaxies (2,401): all galaxies identified by {\sc hdbscan} to lie in groups or clusters. For simplicity, hereafter we refer to all the 
 galaxies in these environments as cluster galaxies. 
 \item Filament galaxies (766): all galaxies which are within a radius of 1 Mpc of the filamentary spine (see Sec.~\ref{web} for details on the choice 
 of filament radius) detected by \disperse.
 \item Void galaxies (1,113): all other galaxies which lie within $170^\circ \leq \rm{RA} \leq 200^\circ, 17^\circ \leq \rm{Dec} \leq 33^\circ$, 
  and have redshift $0.013 \leq z \leq 0.033$, but are not selected in either of the above two categories.
 \end{itemize}
 Figure~\ref{env} shows the projected distribution of galaxies belonging to various environments in the Coma supercluster.

%%%%%%%%%%%%%%%%%%%%%%%%%%%%%%

 \section{Broadband colours of galaxies in different environments}
 \label{colour}

 In this section we test the impact of local environment (Sec.~\ref{s:env}) on optical and UV colours of galaxies. It is worth
 noting that although such studies have been conducted for various samples of groups and clusters at different redshifts, 
 no other large-scale structure offers such a {\it continuous range of `local' environments} in the nearby Universe 
 \citep[e.g.][]{gavazzi10,mahajan10,cybulski14}. 
 The proximity of the Coma supercluster also makes it a special laboratory to investigate the properties of not just the giant, but also the dwarf
 galaxies. But since \smass~is well correlated with environment \citep[e.g.][]{haines07}, and may be the underlying property governing other galaxy 
 observables \citep{kauff03}, where applicable we perform the following analysis in three classes of environments and two bins of luminosity. 
 For the latter we divide our sample into dwarfs ($z\geq15$ mag) and giants ($z<15$) following \citet{mahajan10}. At the redshift of the Coma cluster
 ($z=0.023$), $z=15$ mag corresponds to a $M^*\sim 10^{9.5} M_{\odot}$.
 
 \begin{figure}
 \centering{
 {\rotatebox{270}{\epsfig{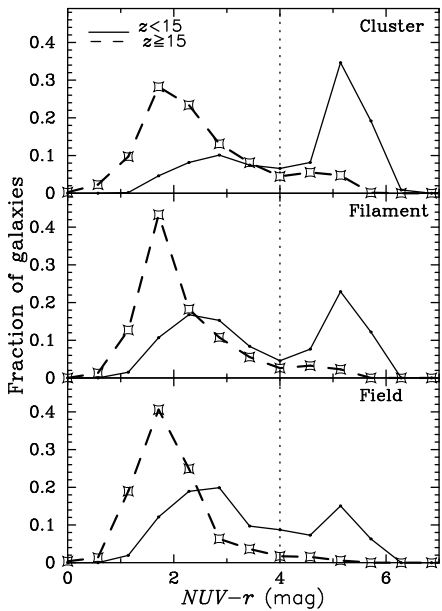}}}}
 \caption{The \nuvr~colour distribution of dwarf ($z\geq15$) and giant ($z<15$) galaxies in three different environments in the
 Coma supercluster.}
 \label{nuvr}
 \end{figure}

 Figure~\ref{nuvr} shows the distribution of the \nuvr~colour for the dwarf and giant galaxies in the three different
 environments in the Coma supercluster. As expected, the fraction of red giants declines, while that of the dwarfs increases as the density of
 environment reduces. The dwarfs are almost always blue except in the densely populated clusters and groups, although in much fewer 
 numbers than their giant counterparts. \citet[][also see \citet{schawinski07}]{wyder07} found the \nuvr~distribution of SDSS galaxies 
 to be bimodal around \nuvr~$\sim 4$ mag. Using their criteria we find 38\%, 20\% and 
 13\% of all UV-detected galaxies in clusters, filaments and voids are passive (i.e. \nuvr~$> 4$ mag) in our sample, respectively. 
 
  \begin{figure}
 \centering{
 {\rotatebox{270}{\epsfig{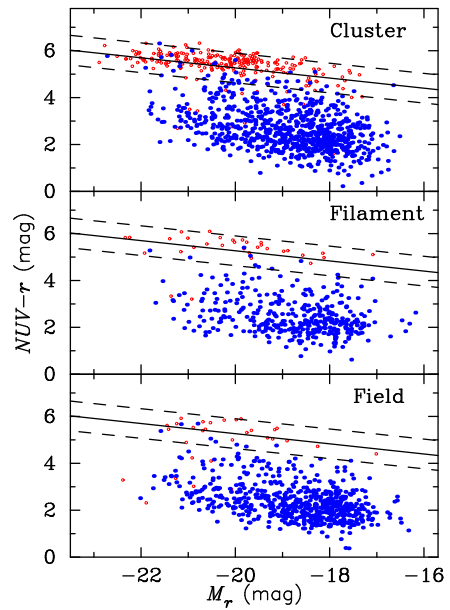}}}}
 \caption{The \nuvr~versus $M_r$ colour-magnitude diagram for the galaxies in three different environments in the Coma supercluster. The points are 
 colour-coded as: passive galaxies (EW(H$\alpha$) $\leq 2$ \AA) as {\it open red 
 symbols}, and star-forming galaxies  (EW(H$\alpha$) $> 2$ \AA) as {\it blue points}. The red sequence fitted to the passive galaxies in clusters
 are repeated in the other two panels.  }
 \label{cmd}
 \end{figure}

 Fig.~\ref{cmd} shows the distribution of non-AGN galaxies in the \nuvr~versus $M_r$ colour-magnitude space for the three different environments. 
 Here we divide the sample into passive and star-forming galaxies around H$\alpha$ equivalent width (EW) $= 2$ \AA\ \citep[e.g.][]{haines08}. 
 The colour-magnitude relation fitted to passive galaxies in clusters takes the form $NUV-r = 0.94 - 0.22 M_r$, with a dispersion of $0.63$ mag. 
 This figure 
 evidently shows the strengthening of the red sequence as the environment becomes denser. We note that in all environments galaxies with 
 \nuvr~$\lesssim 4$ mag are almost always spectroscopically star-forming. It is also notable that despite a small population of passive galaxies in
 the filament region, the distinction between the red sequence and the blue cloud is much clearly visible in the filaments relative to the void sample. 
 
 These observations are in agreement with the results of \citet{gavazzi10}, who %showed that at  \citet{gavazzi10} 
 made use of the $M_i$ vs $g-i$ colour-magnitude relation to show that the red sequence becomes more apparent and the fraction of blue 
 galaxies declines with increasing environmental density.
 Similar trends have also been observed in and around the Virgo cluster, where using the
 $NUV-i$ colour \citet{boselli14} showed that the red sequence is well established in the low density outskirts of the cluster, albeit with a much lower
 fraction of red galaxies relative to their star-forming counterparts.

 Overall, in agreement with the literature \citep[e.g.][]{wyder07,haines08} Fig.~\ref{cmd} shows that the position of a galaxy in the 
 \nuvr~versus $M_r$ colour-magnitude space can be robustly used to distinguish
 the star-forming galaxies from their passively-evolving counterparts.

 \begin{figure}
 \centering{
 {\rotatebox{270}{\epsfig{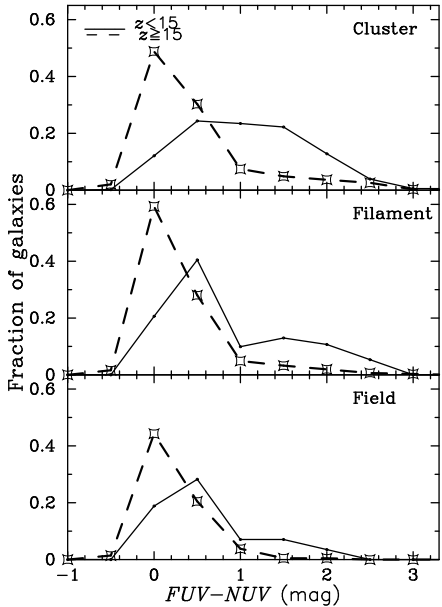}}}}
 \caption{The \fnuv~colour distribution of dwarf ($z\geq15$) and giant ($z<15$) galaxies in three different environments in the
 Coma supercluster.}
 \label{fnuv}
 \end{figure}
 
 The trends seen for \nuvr~ also continue in the distribution of the \fnuv~distributions in different environments as shown in Figure~\ref{fnuv}. 
 The UV colour is a good proxy for young massive stars formed recently in star-forming galaxies. In the passively-evolving galaxies the $FUV$ 
 emission is dominated by evolved stellar populations, and hence correlates well with both age and metallicity
 of the dominating stellar population with surprisingly little scatter \citep{burstein88,oconnell99,boselli05b,smith12}. 
 Therefore as expected, the fraction of massive galaxies having \fnuv~$\gtrsim 1$ mag 
 increases from 27\% in the field to 63\% in the clusters with the filaments bridging the gap in between with the fraction of red galaxies $= 39\%$.
 The dwarfs on the other hand are relatively bluer than their giant counterparts in all environments, with the fraction of galaxies with 
 \fnuv~$\gtrsim 1$ mag being 19\%, 11\% and 7\% in the clusters, filament and field region respectively. Segregating galaxies by their SF activity
 based on EW($H\alpha$), \citet{cortese05} also observed similar trends for the galaxies in Abell~1367 such that the star-forming galaxies
 dominate at high UV luminosities, while the quiescent ones contribute at the faint end of the UV luminosity function.

  \begin{figure}
 \centering{
 {\rotatebox{270}{\epsfig{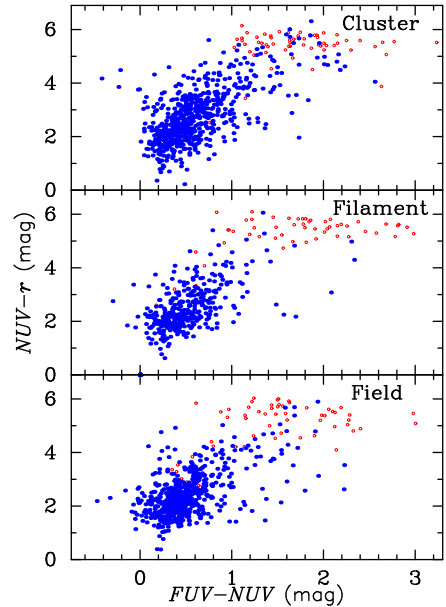}}}}
 \caption{The \nuvr~versus \fnuv~colour-colour distribution of star-forming and passively evolving galaxies in three different environments in the
 Coma supercluster.}
 \label{uv-col}
 \end{figure}
 
 All galaxies detected in UV are shown in the \nuvr~versus \fnuv~colour-colour distribution in Fig.~\ref{uv-col}. While \nuvr~and \fnuv~become redder
 together for the blue galaxies (EW(H$\alpha$)$\geq2$\AA), the \nuvr~is always between 5--6 mag for passively-evolving galaxies independent 
 of their \fnuv~colour. Therefore, just like Fig.~\ref{cmd}, even the position of a UV-detected galaxy in the UV colour-colour space can be robustly
 used to distinguish star-forming galaxies from those which are evolving passively \citep[e.g.][]{gildepaz07}.

 In a nutshell, in agreement with other studies of the Coma supercluster \citep{gavazzi10,mahajan10} this section shows that not just the hostile 
 environment of clusters, but all components of the cosmic-web influence the broadband colours of the dwarf and giant galaxies in the Coma supercluster. 

 %%%%%%%%%%%%%%%%%%%%%%%%%%
  \section{Properties of galaxies as a function of distance from clusters and spine of filaments}
  \label{web}
  
 \begin{figure}
 \centering{
 {\rotatebox{270}{\epsfig{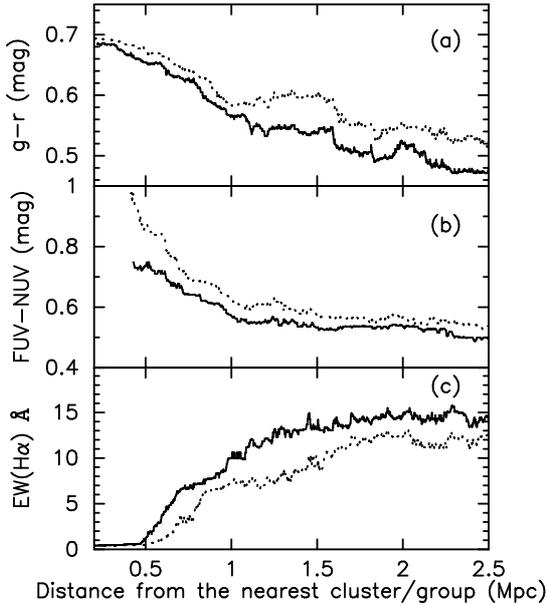}}}
 \caption{ The (a) $g-r$ colour, (b) \fnuv~colour and, (c) median EW(H$\alpha$) of all {\it (dotted line)} and non-AGN 
 {\it (solid line)} galaxies as a function of their distance from the nearest cluster or group. These trends show the expected trend that 
 galaxies progressively become bluer and star-forming further away from the centre of clusters.}
 \label{clusdist}}
 \end{figure}

 The observed trend of declining SFR with decreasing cluster-centric distance has been well explored with different datasets in the literature
 \citep[e.g.][]{lewis02,gomez03,balogh04}. Here we reconfirm this trend in the Coma supercluster by analysing 
 the optical and UV properties of galaxies % we now analyse properties of galaxies 
 as a function of their distance from the centre of the nearest cluster or group. We extend this analysis a step further by also 
 investigating the impact of the cosmic-web on the properties of galaxies as a function of their distance from the spine of the filaments.  
 In Fig.~\ref{clusdist} we show the median EW(H$\alpha$), \fnuv~and the $g-r$ colour of all as well as non-AGN galaxies as a function of the 
 distance from the nearest group or cluster (Sec.~\ref{groups}). It is notable that excluding AGN from the sample
 only changes the normalization of the radial distribution of galaxies, not the observed trends. Therefore, in the following we discuss the 
 complete sample of galaxies including AGN.
 
 It is evident that the density of environment has an impact on the colours as well as the EW(H$\alpha$) of galaxies, 
 such that the median \fnuv, as well as the $g-r$ colour of galaxies progressively become bluer, and their emission in H$\alpha$ increases
 away from the cluster centre. Beyond $\sim 1.5$ Mpc from the cluster centre, both colours and 
 EW(H$\alpha$) seem to approach a constant 
 value close to the respective medians for the void galaxies. Therefore, in agreement with the literature  \citep{gavazzi10,boselli14}
 Fig.~\ref{clusdist} shows that the fraction of star-forming, blue galaxies in the Coma supercluster increases with the cluster-centric distance. 
 % viz. $g-r = 0.42$ mag and EW(H$\alpha$) $= 15.5$ \AA.

 \begin{figure}
 \centering{
 {\rotatebox{270}{\epsfig{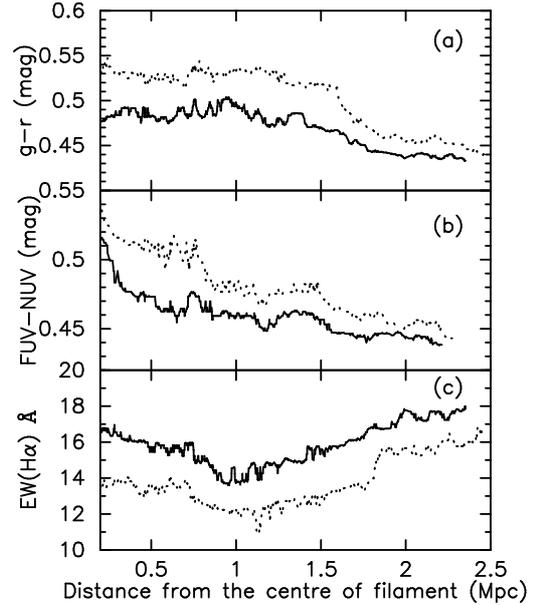}}}
 \caption{The (a) $g-r$ colour, (b) \fnuv~colour and, (c) median EW(H$\alpha$) of all {\it (dotted line)} and non-AGN 
 {\it (solid line)} galaxies in the Coma supercluster as a function of their distance from the spine of the filament. These trends evidently 
 show that some quenching mechanism comes to play as galaxies close in from voids to the spine of filaments. }
 \label{spine}}
 \end{figure}
 
 In Fig.~\ref{spine} we show the median EW(H$\alpha$), \fnuv~and the $g-r$ colour of all as well as non-AGN galaxies in the Coma 
 supercluster as a function of their distance from the spine of the filament (Sec.~\ref{s:env}).  
 While both the colours are constant for galaxies within $1$ Mpc of the spine,
 they decline by $\gtrsim 0.7$ mag beyond it. The trend seen for colours is replicated for the EW(H$\alpha$) distribution, which is 
 $\sim 13$\AA\ within 1 Mpc of the spine, but increases by another $\sim 3$\AA\ beyond 1 Mpc from it. 
 We tested the statistical significance of these trends by dividing the distributions for all the quantities, $g-r$, \fnuv~and EW(H$\alpha$)
 into three radial bins: $\leq r_{200}$\footnote{$r_{200}$ is defined as the radius of the sphere centred
  at the cluster centre within which the matter density is 200 times the critical density of the Universe.}, 
  $1-2r_{200}$ and $>2 r_{200}$, respectively. The Kolmogorov-Smirnov (K-S) statistic confirms that the
 probability for the EW(H$\alpha$) distributions in the first and third radial bin to have been drawn from the same parent sample is 
 $2.68e-05$, while the same for the $g-r$ and \fnuv~colour is $9.76e-08$ and $5.73e-14$, respectively. 
 The same for non-AGN galaxies is $3.34e-04, 1.61e-07$ and $3.07e-10$, respectively.
 
% the difference between the c.f. is =  0.154370442     : K-S output for Halpha
% probability of null hypothesis =   2.68131662E-05
% the difference between the c.f. is =  0.143081546     : non-AGN only
% probability of null hypothesis =   3.33716423E-04
%
% the difference between the c.f. is =  0.189059079     : for g-r
% probability of null hypothesis =   9.76236905E-08
% the difference between the c.f. is =  0.196076751     : non-AGN only
%probability of null hypothesis =   1.60980051E-07
%
% the difference between the c.f. is =  0.337347895     : for FUV-NUV
% probability of null hypothesis =   5.73525195E-14
% the difference between the c.f. is =  0.275523394     : non-AGN only
% probability of null hypothesis =   3.07528031E-10

 Based on these results we choose to limit the radius of filaments in this work to 1 Mpc as mentioned in Sec.~\ref{ss:env}. This is also in agreement
 with other studies of large-scale filaments based on the SDSS data \citep{kuutma17}. A caveat in this analysis however is that the SDSS spectra
 taken with a $3^{\prime\prime}$ diameter fibre only represents the central bulge of large galaxies. Hence it underestimates the effects of quenching
 due to various environmental mechanisms primarily occurring in the outer regions of large galaxies \citep{koopmann06,gavazzi13}. 

 Together, Figs.~\ref{clusdist} and \ref{spine} show that galaxies are not just transformed when they assimilate into dense clusters and groups,
 but also as they approach the seemingly less hostile environment of the filaments. We discuss the implications of this result in the context of the
 existing literature in the following section.

 \section{Discussion}
 \label{discuss}

 \subsection{The significance of large-scale environment for galaxy evolution}
 
 In agreement with the literature our results show that the large-scale filaments are critical to transformations in a galaxy prior to its assimilation into
 clusters or groups. With WHIM temperatures of $3-8$ keV and hot gas fraction of $\sim 10\%$ \citep[e.g.][]{dietrich12}, 
 filaments are not just passages leading galaxies to their final destination, but offer a unique environment in which the star formation properties
 of galaxies can change dramatically over a short period of time \citep{boue08,fadda08,porter08,mahajan12}. 
 
 Using a sample of straight 
 filaments selected from the Two-degree field galaxy redshift survey, \citet{porter08} showed that the statistical parameter $\eta$ derived from
 spectra and acting as a proxy for the SFR, is enhanced between 2--3 Mpc from the edge of the filaments. Moreover, using a sample of 
 Abell clusters ($z\leq0.12$)
 selected from the SDSS, \citet{mahajan12} found that the SFR of galaxies is enhanced on the outskirts of clusters, especially the ones which
 are fed by more filaments. Elsewhere, such an enhancement in star formation is detected in galaxies on filaments feeding clusters using infrared 
 data \citep{fadda08}. These results are intriguing given that most of the star formation in supercluster galaxies at low redshift is contributed by 
 normal galaxies on the periphery of clusters \citep{haines11}. 

 At high redshift ($z>0.2$), several authors have made use of multi-wavelength data to investigate large-scale filaments. For instance,
 \citet{geach11} combined the $NUV$ data from the \g~with the 24-$\mu$m data from the {\it Spitzer} space telescope to 
 investigate a panoramic 15-Mpc region around the supercluster Cl 0016+16 ($z=0.55$). In agreement with studies at $z \lesssim 0.1$,
 their results \citep[also see][]{coppin12,darvish14,darvish15} suggest that the star formation rate of 
 galaxies may increase before they fall into clusters. It has also been suggested that the infalling galaxies may experience a
 phase of obscured star formation before being effected by the cluster environment \citep{gallazzi09}. 
   
 Owing to their highly coherent motion along linear filaments \citep{gonzalez09}, infalling galaxies are more likely to interact with each other 
 gravitationally relative to their counterparts in voids. Such fly-by interactions can trigger burst of star formation in galaxies falling along 
 filaments \citep{porter08,mahajan12}. But such an enhancement in SFR of galaxies will also increase their star formation efficiency, hence 
 assuming no new infall of gas, such galaxies can become redder relative to similar void galaxies even before encountering the cluster 
 environment. The observations presented in this work as well as other studies \citep{alpaslan16,kuutma17,kraljic18} support this hypothesis.
 
 Along with the trend in galaxy properties as a function of cluster-centric distance, now there is also a growing consensus that the properties
 of galaxies on filaments change as a function of their distance from the backbone of the
 cosmic-web. Using data from the SDSS, \citet{kuutma17} found a higher elliptical-to-spiral ratio, increasing $g-i$ colour and decreasing 
 SFR for galaxies moving from voids to filaments. In fact they found the latter trends in the $g-i$ colour and SFR to be persistent even after 
 separating spiral and elliptical galaxies ($M_r\leq-20$ mag). 
 Using UV-derived SFR for a sample of $\sim 12,000$ galaxies ($z<0.09$) from the Galaxy and Mass Assembly (GAMA) survey,
 complete to a stellar mass of $\sim 10^9 M_{\odot}$, \citet{alpaslan16} showed that the galaxies at the core of filaments are more massive and 
 have lower specific SFR (sSFR; SFR/$M^*$) relative to their counterparts at the edges of voids. By using a complementary dataset from the 
 GAMA survey, \citet{kraljic18} also reached similar conclusions. These results are in agreement with the
 comparative analysis of properties of galaxies falling into groups isotropically and along filaments. In this comparison, \citet{martinez16} found that the 
 luminosity function of galaxies in filaments is indistinguishable from the galaxies infalling isotropically, yet their sSFR is lower than the latter,
 suggesting that galaxies in filaments are more strongly quenched. 
 
 Recently \citet{kleiner17} examined the HI-to-$M^*$ (HI fraction) of galaxies in the nearby ($< 181$ Mpc) Universe using data from the 
 6-degree Field Galaxy Survey and stacking HI data from the HI Parkes all sky survey. They found that while the HI fraction for galaxies 
 with $M^*<10^{11} M_{\odot}$ is the same on filaments and in the control sample, vice versa is true for galaxies more massive 
 than $10^{11} M_{\odot}$. \citet{kleiner17} suggested that this result is evidence of cold mode accretion of gas in massive galaxies 
 on filaments, owing to their larger potential well. Our results are in broad agreement with \citet{kleiner17} such that in this work we have shown 
 that the intermediate density environment prevalent in the filaments effect dwarfs as well as the giants, although this effect is observed to be 
 more pronounced in the latter. But since we do not find enough galaxies observed in HI for the entire Coma supercluster, in this work we can 
 neither support nor reject the hypothesis of cold mode accretion proposed by \citet{kleiner17}. 
  
 The results presented here are also in broad agreement with those of \citet{gavazzi13}. Using samples of HI-rich and HI-poor late-type galaxies 
 (LTGs; $M_{HI}/M_{\odot}\geq9$) in the Local supercluster and part of the Coma supercluster from the ALFALFA survey, they showed
 that HI-rich LTGs represent $\sim 60\%$ of all LTGs in the low and intermediate density environments, and drop to zero at the centre of the 
 Coma cluster. On the other hand, the frequency of HI-poor LTGs increases with increasing galaxy density.

 \subsection{How does filaments effect galaxies}
  
 The census of observed baryons in the local Universe falls short of their estimated contribution of 5 per cent to the total energy budget by 
 a factor of two. Cosmological simulations indicate that rather than condensing into virialized haloes, the missing baryons 
 reside in the filamentary cosmic-web. 
 The symbiotic relation between the WHIM ($10^5-10^7$ K), and the constituent galaxies of the filaments is well represented by the discovery of a
 bent double lobe radio source (DLRS) in a filament feeding the rich cluster Abell~1763 \citep{edwards10}. Assuming that the bend in the jet of the
 DLRS is due to the ram-pressure experienced by it 3.4 Mpc away from the centre of the cluster, \citet{edwards10} constrained the density of 
 WHIM to be a few times $10^{-29}$ gm cm$^{-3}$, in agreement with the literature, thereby evidently showing that environmental processes 
 such as ram-pressure stripping may operate much farther away from high density environment of clusters. % and effect galaxies on filaments.
 
 By studying the spectroscopic properties of 28 star-forming galaxies ($z\sim0.53$) in the COSMOS field, \citet{darvish15} found that within 
 uncertainties, the EW, EW versus sSFR relation, EW versus $M^*$ relation, line-of-sight velocity dispersion, dynamical mass, and 
 stellar-to-dynamical mass ratio are similar for filament and field star-forming galaxies. Yet on average, star-forming galaxies on filaments
 are more metal enriched ($0.1-0.5$ dex) relative to their field counterparts. \citet{darvish15} suggest that the high metallicity may have been 
 caused by the inflow of the enriched intrafilamentary gas into galaxies residing on filaments. However, these high redshift observations are in 
 conflict with the results of \citet{hughes13} who found that even though gas-poor galaxies in the Virgo cluster ($z\sim0$) are typically more 
 metal-rich, statistically the $M^*$-metallicity relation is independent of environment. \citet{hughes13} also demonstrated that removal of gas 
 from the outer regions of disc galaxies may increase the observed metallicity by $\sim 0.1$ dex. 

  \citet{darvish15} also showed that electron densities are significantly lower by a factor of $\sim 17$ in filament star-forming galaxies compared 
  to those in the field, possibly because of a longer star-formation timescale for filament star-forming galaxies. 
 This, and other aforementioned studies complement the results found in this work and highlight that galaxy properties are not just shaped by 
 $M^*$ and the large-scale galaxy density, but also the tidal effects of the anisotropic cosmic-web resulting from the interactions between the 
 intra-filamentary gas and galaxies.
 
  \subsection{Epilogue}
 
 Several innovative techniques such as gravitational lensing \citep{van14}, stacking of optical photometric data \citep{zhang13}, and stacking of 
 x-ray data \citep{fraser11} are now being employed to discover and characterise various aspects of the cosmic-web.  
 Therefore, observations such as those presented here are crucial not just for strengthening our models of the formations and evolution of the Universe,
 but also to serve as a benchmark to compare with discoveries and trends observed at higher redshift. 
 In the near future, state-of-the-art technology and global observational projects will play a key role in enhancing our understanding of the cosmic-web.
 To prepare ahead for these data, while some studies are utilizing simulations to examine the
 evolution of the cosmic-web properties with redshift \citep{gheller16}, others are trying to predict the 21-cm signal of WHIM which may be detected
 by the Square Kilometer Array (SKA) \citep[][also see \citet{tejos16}]{horii17}. Elsewhere, x-ray observations of large-scale filaments have been 
 reported \citep{werner08,eckert15}.
 With an ever expanding database of the known large-scale structures at different epochs, we can hope to understand the evolution of the environment 
 and properties of galaxies in filaments in the near future.
 
 \section{Summary}
 \label{results}

 We have utilized the optical data from the SDSS DR 12 and UV data from the \g~survey to present a catalogue of UV detected galaxies in the Coma 
 supercluster. We have used \disperse~to characterize the large-scale filaments, and {\sc hdbscan} algorithm to identify the groups and clusters of 
 galaxies in the Coma supercluster. The key results reported in this work are:
 \begin{itemize}
 \item The position of a galaxy in the $NUV-r$ versus $r$ colour-magnitude space and in the \fnuv~versus $NUV-r$ colour-colour space can be robustly used
 to separate passively-evolving galaxies from their star-forming counterparts in all environments: clusters, filaments and voids.
 \item Dwarf galaxies are almost always blue except in the dense interiors of the clusters and groups. On the other hand, 
 most of the giants are red in all environments in the Coma supercluster, but their fraction decreases with the environmental density. These trends
 are seen for the \fnuv, as well as the $NUV-r$ colour.
 \item The $g-r$ and \fnuv~colour and EW(H$\alpha$) for galaxies vary as a function of their distance from the nearest cluster or group viz.,
 galaxies become redder and emission in H$\alpha$ declines with their distance from the nearest maximum in density.
 \item Within a radius of 1 Mpc from the spine of the filament, galaxies become redder and the median EW(H$\alpha$) declines closer to the spine
 of the filament. This transformation in $g-r$, \fnuv, and H$\alpha$ emission of galaxies is statistically significant, and therefore evidently depicts the
 key role played by the large-scale filaments in shaping properties of galaxies.
 \end{itemize}
 
\section*{Acknowledgements}

 We are grateful to Prof. Jasjeet Singh Bagla for his suggestions and comments on several aspects of this work. 
 We acknowledge the high power computing (HPC) facility at IISER Mohali. We are grateful to the anonymous reviewer whose suggestions 
 helped to elucidate this manuscript. 
Funding for SDSS-III has been provided by the Alfred P. Sloan Foundation, the Participating Institutions, the National Science Foundation, and the U.S. Department of Energy Office of Science. The SDSS-III web site is http://www.sdss3.org/.
SDSS-III is managed by the Astrophysical Research Consortium for the Participating Institutions of the SDSS-III Collaboration including the University of Arizona, the Brazilian Participation Group, Brookhaven National Laboratory, Carnegie Mellon University, University of Florida, the French Participation Group, the German Participation Group, Harvard University, the Instituto de Astrofisica de Canarias, the Michigan State/Notre Dame/JINA Participation Group, Johns Hopkins University, Lawrence Berkeley National Laboratory, Max Planck Institute for Astrophysics, Max Planck Institute for Extraterrestrial Physics, New Mexico State University, New York University, Ohio State University, Pennsylvania State University, University of Portsmouth, Princeton University, the Spanish Participation Group, University of Tokyo, University of Utah, Vanderbilt University, University of Virginia, University of Washington, and Yale University. 
This research made use of {\sc topcat} \citep{taylor05} and the ``K-corrections calculator'' service available at http://kcor.sai.msu.ru/.

 Mahajan is funded by the INSPIRE Faculty award (DST/INSPIRE/04/2015/002311), Department of Science and Technology (DST), 
 Government of India. Shobhana was supported by the INSPIRE scholarship for higher education for her BS-MS dual degree at IISERM.

%%%%%%%%%%%%%%%%%%%%%%%%%%%%%%%%%%%%%%%%%%%%%%%%%%
%%%%%%%%%%%%%%%%%%%% REFERENCES %%%%%%%%%%%%%%%%%%%%%%

%%%%%%%%%%%%%%%%%%%%%%%%%%%%%%%%%%%%%%%%%%%%%%%%%%

% Don't change these lines
\bsp	% typesetting comment
\label{lastpage}


\begin{thebibliography}{99}

 \bibitem[\protect\citeauthoryear{Akamatsu et al.}{2017}]{akamatsu17} Akamatsu H., et al., 2017, A\&A, 606, A1 
 \bibitem[\protect\citeauthoryear{Alam et al.}{2015}]{alam15} Alam S., et al., 2015, ApJS, 219, 12 
 \bibitem[\protect\citeauthoryear{Alpaslan et al.}{2016}]{alpaslan16} Alpaslan M., et al., 2016, MNRAS, 457, 2287 

 \bibitem[\protect\citeauthoryear{Baldwin, Phillips, \& Terlevich}{1981}]{bpt} Baldwin J.~A., Phillips M.~M.,
  Terlevich R., 1981, PASP, 93, 5 
  \bibitem[\protect\citeauthoryear{Balogh et al.}{2004}]{balogh04} Balogh M., et al., 2004, MNRAS, 348, 1355 
  \bibitem[\protect\citeauthoryear{Bernstein et al.}{1995}]{bernstein95} Bernstein G.~M., Nichol R.~C., Tyson J.~A., Ulmer M.~P., 
  Wittman D., 1995, AJ, 110, 1507 
 \bibitem[\protect\citeauthoryear{Bianchi et al.}{2011}]{bianchi11} Bianchi L., Herald J., Efremova B., Girardi L., Zabot A.,
 Marigo P., Conti A., Shiao B., 2011, Ap\&SS, 335, 161 
 \bibitem[\protect\citeauthoryear{Bianchi, Conti, \& Shiao}{2014}]{bianchi14} Bianchi L., Conti A., Shiao B., 2014, AdSpR, 53, 900 
 \bibitem[\protect\citeauthoryear{Biviano et al.}{2011}]{biviano11} Biviano A., Fadda D., Durret F., Edwards L.~O.~V., Marleau F., 
 2011, A\&A, 532, A77 
 \bibitem[\protect\citeauthoryear{Bonjean et al.}{2018}]{bonjean18} Bonjean V., Aghanim N., Salom{\'e} P., Douspis M., 
 Beelen A., 2018, A\&A, 609, A49 
% \bibitem[\protect\citeauthoryear{Boselli et al.}{2005a}]{boselli05a} Boselli A., et al., 2005, ApJ, 623, L13 
 \bibitem[\protect\citeauthoryear{Boselli et al.}{2005b}]{boselli05b} Boselli A., et al., 2005, ApJ, 629, L29 
 \bibitem[\protect\citeauthoryear{Boselli et al.}{2014}]{boselli14} Boselli A., et al., 2014, A\&A, 570, A69 
 \bibitem[\protect\citeauthoryear{Bou{\'e} et al.}{2008}]{boue08} Bou{\'e} G., Durret F., Adami C., Mamon G.~A., Ilbert O., Cayatte V.,
 2008, A\&A, 489, 11 
 \bibitem[\protect\citeauthoryear{Budav{\'a}ri et al.}{2009}]{budavari09} Budav{\'a}ri T., et al., 2009, ApJ, 694, 1281  
 \bibitem[\protect\citeauthoryear{Burstein et al.}{1988}]{burstein88} Burstein D., Bertola F., Buson L.~M., Faber S.~M., Lauer T.~R., 
 1988, ApJ, 328, 440 

 \bibitem[\protect\citeauthoryear{Campello et al.}{2013}]{campello13} Campello, R.~J.~G.~B., Moulavi, D., Zimek, A., Sander, J., 2013, 
 Data mining and knowledge discovery, Vol. 27, Issue 3, 2013, p 344
 \bibitem[\protect\citeauthoryear{Campello et al.}{2015}]{campello15} Campello, R.~J.~G.~B., Moulavi, D., Zimek, A., Sander, J., 2015, 
 ACM Trans. Knowl. Discov. Data, 10, 51
 \bibitem[\protect\citeauthoryear{Cardelli, Clayton, \& Mathis}{1989}]{cardelli89} Cardelli J.~A., Clayton G.~C., Mathis J.~S., 1989, ApJ, 345, 245 
 \bibitem[\protect\citeauthoryear{Cautun \& van de Weygaert}{2011}]{cautun11} Cautun M.~C., van de Weygaert R., 
 2011, ascl.soft, ascl:1105.003 
 \bibitem[\protect\citeauthoryear{Chilingarian \& Zolotukhin}{2012}]{chilingarian12} Chilingarian I.~V., Zolotukhin I.~Y., 2012,
 MNRAS, 419, 1727 
 \bibitem[\protect\citeauthoryear{Chincarini \& Rood}{1976}]{chincarini76} Chincarini G., Rood H.~J., 1976, ApJ, 206, 30
 \bibitem[\protect\citeauthoryear{Coppin et al.}{2011}]{coppin11} Coppin K.~E.~K., et al., 2011, MNRAS, 416, 680  
 \bibitem[\protect\citeauthoryear{Coppin et al.}{2012}]{coppin12} Coppin K.~E.~K., et al., 2012, ApJ, 749, L43 
 \bibitem[\protect\citeauthoryear{Cornett et al.}{1998}]{cornett98} Cornett R.~H., et al., 1998, AJ, 116, 44 
 \bibitem[\protect\citeauthoryear{Cortese et al.}{2005}]{cortese05} Cortese L., et al., 2005, ApJ, 623, L17 
 \bibitem[\protect\citeauthoryear{Cortese et al.}{2007}]{cortese07} Cortese L., et al., 2007, MNRAS, 376, 157 
% \bibitem[\protect\citeauthoryear{Cortese, Gavazzi, \& Boselli}{2008}]{cortese08} Cortese L., Gavazzi G., Boselli A., 2008, MNRAS, 390, 1282 
% \bibitem[\protect\citeauthoryear{Cortese et al.}{2012}]{cortese12} Cortese L., et al., 2012, A\&A, 544, A101 
 \bibitem[\protect\citeauthoryear{Cybulski et al.}{2014}]{cybulski14} Cybulski R., Yun M.~S., Fazio G.~G., Gutermuth R.~A.,
 2014, MNRAS, 439, 3564 

 \bibitem[\protect\citeauthoryear{Darvish et al.}{2014}]{darvish14} Darvish B., Sobral D., Mobasher B., Scoville N.~Z., Best P.,
  Sales L.~V., Smail I., 2014, ApJ, 796, 51 
 \bibitem[\protect\citeauthoryear{Darvish et al.}{2015}]{darvish15} Darvish B., Mobasher B., Sobral D., Hemmati S., Nayyeri H., 
 Shivaei I., 2015, ApJ, 814, 84 
 \bibitem[\protect\citeauthoryear{Dietrich et al.}{2012}]{dietrich12} Dietrich J.~P., Werner N., Clowe D., Finoguenov A., Kitching T., Miller L.,
  Simionescu A., 2012, Natur, 487, 202 
 \bibitem[\protect\citeauthoryear{Donas et al.}{1990}]{donas90} Donas J., Milliard B., Laget M., Buat V., 1990, A\&A, 235, 60 
 \bibitem[\protect\citeauthoryear{Donas, Milliard, \& Laget}{1995}]{donas95} Donas J., Milliard B., Laget M., 1995, A\&A, 303, 661 
 \bibitem[\protect\citeauthoryear{Dressler}{1980}]{dressler80} Dressler A., 1980, ApJ, 236, 351 
 
 \bibitem[\protect\citeauthoryear{Eckert et al.}{2015}]{eckert15} Eckert D., et al., 2015, Natur, 528, 105 
 \bibitem[\protect\citeauthoryear{Edwards et al.}{2010a}]{edwards10} Edwards L.~O.~V., Fadda D., Frayer D.~T., Lima Neto G.~B., Durret F.,
  2010, AJ, 140, 1891 
  \bibitem[\protect\citeauthoryear{Edwards, Fadda, \& Frayer}{2010b}]{edwards10b} Edwards L.~O.~V., Fadda D., Frayer D.~T., 2010, ApJ, 724, L143 
 \bibitem[\protect\citeauthoryear{Ester et al.}{1996}]{ester96} Ester M., Kriegel H.-P., Sander J., Xu X., 1996, AAAI press, 226
 
 \bibitem[\protect\citeauthoryear{Fadda et al.}{2008}]{fadda08} Fadda D., Biviano A., Marleau F.~R., Storrie-Lombardi L.~J., Durret F.,
  2008, ApJ, 672, L9 
 \bibitem[\protect\citeauthoryear{Fontanelli}{1984}]{fontanelli84} Fontanelli P., 1984, A\&A, 138, 85 
 \bibitem[\protect\citeauthoryear{Fraser-McKelvie, Pimbblet, \& Lazendic}{2011}]{fraser11} Fraser-McKelvie A., Pimbblet K.~A.,
  Lazendic J.~S., 2011, MNRAS, 415, 1961 

 \bibitem[\protect\citeauthoryear{Gallazzi et al.}{2009}]{gallazzi09} Gallazzi A., et al., 2009, ApJ, 690, 1883 
 \bibitem[\protect\citeauthoryear{Gavazzi et al.}{2010}]{gavazzi10} Gavazzi G., Fumagalli M., Cucciati O., Boselli A., 2010, A\&A, 517, A73 
% \bibitem[\protect\citeauthoryear{Gavazzi, Savorgnan, \& Fumagalli}{2011}]{gavazzi11} Gavazzi G., Savorgnan G., Fumagalli M., 2011, A\&A, 534, A31 
 \bibitem[\protect\citeauthoryear{Gavazzi et al.}{2013}]{gavazzi13} Gavazzi G., et al., 2013, A\&A, 553, A90 
% \bibitem[\protect\citeauthoryear{Gavazzi et al.}{2015}]{gavazzi15} Gavazzi G., et al., 2015, A\&A, 576, A16 
 \bibitem[\protect\citeauthoryear{Geach et al.}{2011}]{geach11} Geach J.~E., Ellis R.~S., Smail I., Rawle T.~D., Moran S.~M., 2011, 
 MNRAS, 413, 177 
 \bibitem[\protect\citeauthoryear{G{\'e}nova-Santos et al.}{2015}]{genova15} G{\'e}nova-Santos R., Atrio-Barandela F., Kitaura F.-S.,
  M{\"u}cket J.~P., 2015, ApJ, 806, 113 
  \bibitem[\protect\citeauthoryear{Gheller et al.}{2016}]{gheller16} Gheller C., Vazza F., Br{\"u}ggen M., Alpaslan M., Holwerda B.~W., 
  Hopkins A.~M., Liske J., 2016, MNRAS, 462, 448 
  \bibitem[\protect\citeauthoryear{Gil de Paz et al.}{2007}]{gildepaz07} Gil de Paz A., et al., 2007, ApJS, 173, 185 
  \bibitem[\protect\citeauthoryear{G{\'o}mez et al.}{2003}]{gomez03} G{\'o}mez P.~L., et al., 2003, ApJ, 584, 210 
 \bibitem[\protect\citeauthoryear{Gonz{\'a}lez \& Padilla}{2009}]{gonzalez09} Gonz{\'a}lez R.~E., Padilla N.~D., 2009, MNRAS, 397, 1498 
 \bibitem[\protect\citeauthoryear{Gregory \& Thompson}{1978}]{gregory78} Gregory S.~A., Thompson L.~A., 1978, ApJ, 222, 784 

 \bibitem[\protect\citeauthoryear{Haines et al.}{2007}]{haines07} Haines C.~P., Gargiulo A., La Barbera F., Mercurio A., Merluzzi P., 
 Busarello G., 2007, MNRAS, 381, 7 
 \bibitem[\protect\citeauthoryear{Haines, Gargiulo, \& Merluzzi}{2008}]{haines08} Haines C.~P., Gargiulo A., Merluzzi P., 2008, MNRAS, 385, 1201
 \bibitem[\protect\citeauthoryear{Haines et al.}{2011}]{haines11} Haines C.~P., Busarello G., Merluzzi P., Smith R.~J., Raychaudhury S.,
 Mercurio A., Smith G.~P., 2011, MNRAS, 412, 145  
 \bibitem[\protect\citeauthoryear{Hammer et al.}{2010}]{hammer10} Hammer D., Hornschemeier A.~E., Mobasher B., Miller N., Smith R., Arnouts S., Milliard B., Jenkins L., 2010, ApJS, 190, 43 
 \bibitem[\protect\citeauthoryear{Hammer et al.}{2012}]{hammer12} Hammer D.~M., Hornschemeier A.~E., Salim S., Smith R., Jenkins L., 
 Mobasher B., Miller N., Ferguson H., 2012, ApJ, 745, 177 
 \bibitem[\protect\citeauthoryear{Hicks \& Mushotzky}{2005}]{hicks05} Hicks A.~K., Mushotzky R., 2005, ApJ, 635, L9 
  \bibitem[\protect\citeauthoryear{Hogg et al.}{2004}]{hogg04} Hogg, D. W., Blanton, M. R., Brinchmann, J., et al. 2004, ApJ, 601, L29
 \bibitem[\protect\citeauthoryear{Horii et al.}{2017}]{horii17} Horii T., Asaba S., Hasegawa K., Tashiro H., 2017, PASJ, 69, 73 
 \bibitem[\protect\citeauthoryear{Hughes et al.}{2013}]{hughes13} Hughes T.~M., Cortese L., Boselli A., Gavazzi G., Davies J.~I., 2013, 
 A\&A, 550, A115 

 \bibitem[\protect\citeauthoryear{Jauzac et al.}{2012}]{jauzac12} Jauzac M., et al., 2012, MNRAS, 426, 3369 

 \bibitem[\protect\citeauthoryear{Kauffmann et al.}{2003}]{kauff03} Kauffmann G., et al., 2003, MNRAS, 341, 54 
 \bibitem[\protect\citeauthoryear{Kennicutt}{1998}]{kennicutt98} Kennicutt R.~C., Jr., 1998, ARA\&A, 36, 189 
 \bibitem[\protect\citeauthoryear{Kewley et al.}{2001}]{kewley01} Kewley L.~J., Dopita M.~A., Sutherland R.~S., 
 Heisler C.~A., Trevena J., 2001, ApJ, 556, 121 
 \bibitem[\protect\citeauthoryear{Kim et al.}{1989}]{kim89} Kim K.-T., Kronberg P.~P., Giovannini G., Venturi T., 1989, Natur, 341, 720 
 \bibitem[\protect\citeauthoryear{Kim et al.}{2016}]{kim16} Kim S., et al., 2016, ApJ, 833, 207 
 \bibitem[\protect\citeauthoryear{Kleiner et al.}{2017}]{kleiner17} Kleiner D., Pimbblet K.~A., Jones D.~H., Koribalski B.~S., Serra P., 2017, 
 MNRAS, 466, 4692 
 \bibitem[\protect\citeauthoryear{Koopmann, Haynes, \& Catinella}{2006}]{koopmann06} Koopmann R.~A., Haynes M.~P., Catinella B., 
 2006, AJ, 131, 716 
 \bibitem[\protect\citeauthoryear{Kraljic et al.}{2018}]{kraljic18} Kraljic K., et al., 2018, MNRAS, 474, 547 
 \bibitem[\protect\citeauthoryear{Kuutma, Tamm, \& Tempel}{2017}]{kuutma17} Kuutma T., Tamm A., Tempel E., 2017, A\&A, 600, L6 

 \bibitem[\protect\citeauthoryear{Lewis et al.}{2002}]{lewis02} Lewis I., et al., 2002, MNRAS, 334, 673 

 \bibitem[\protect\citeauthoryear{Mahajan, Haines, \& Raychaudhury}{2010}]{mahajan10} Mahajan S., Haines C.~P.,
  Raychaudhury S., 2010, MNRAS, 404, 1745 
 \bibitem[\protect\citeauthoryear{Mahajan, Haines, \& Raychaudhury}{2011}]{mahajan11} Mahajan S., Haines C.~P., Raychaudhury S., 2011, 
 MNRAS, 412, 1098 
 \bibitem[\protect\citeauthoryear{Mahajan, Raychaudhury, \& Pimbblet}{2012}]{mahajan12} Mahajan S., Raychaudhury S., Pimbblet K.~A.,
  2012, MNRAS, 427, 1252 
  \bibitem[\protect\citeauthoryear{Mahajan}{2013}]{mahajan13} Mahajan S., 2013, MNRAS, 431, L117 
 \bibitem[\protect\citeauthoryear{Martin et al.}{2005}]{martin05} Martin D.~C., et al., 2005, ApJ, 619, L1 
 \bibitem[\protect\citeauthoryear{Mart{\'{\i}}nez, Muriel, \& Coenda}{2016}]{martinez16} Mart{\'{\i}}nez H.~J., Muriel H., Coenda V., 2016,
  MNRAS, 455, 127 
  \bibitem[\protect\citeauthoryear{Mobasher et al.}{2003}]{mobasher03} Mobasher B., et al., 2003, ApJ, 587, 605 
 \bibitem[\protect\citeauthoryear{McInnes \& Healy}{2017}]{mcinnes17} McInnes L., Healy J., 2017, IEEE International Conference
  on Data Mining Workshops (ICDMW), p33
 \bibitem[\protect\citeauthoryear{McInnes, Healy \& Astels}{2017b}]{mcinnes17b} McInnes L., Healy J., Astels, S., 2017b, Journal of 
 Open Source Software, 2(11), 205
 \bibitem[\protect\citeauthoryear{Muldrew et al.}{2012}]{muldrew12} Muldrew S.~I., et al., 2012, MNRAS, 419, 2670 

 \bibitem[\protect\citeauthoryear{O'Connell}{1999}]{oconnell99} O'Connell R.~W., 1999, ARA\&A, 37, 603 

 \bibitem[\protect\citeauthoryear{Parekh et al.}{2017}]{parekh17} Parekh V., Durret F., Padmanabh P., Pandge M.~B., 2017, MNRAS, 470,
  3742 
% \bibitem[\protect\citeauthoryear{Pontzen et al.}{2017}]{pontzen17} Pontzen A., Tremmel M., Roth N., Peiris H.~V., Saintonge A., 
% Volonteri M., Quinn T., Governato F., 2017, MNRAS, 465, 547 
 \bibitem[\protect\citeauthoryear{Porter \& Raychaudhury}{2007}]{porter07} Porter S.~C., Raychaudhury S., 2007, MNRAS, 375, 1409 
\bibitem[\protect\citeauthoryear{Porter et al.}{2008}]{porter08} Porter S.~C., Raychaudhury S., Pimbblet K.~A., Drinkwater M.~J.,
 2008, MNRAS, 388, 1152 
 
% \bibitem[\protect\citeauthoryear{Rasmussen et al.}{2012}]{rasmussen12} Rasmussen J., Mulchaey J.~S., Bai L., Ponman T.~J., 
% Raychaudhury S., Dariush A., 2012, ApJ, 757, 122 
 \bibitem[\protect\citeauthoryear{Rines et al.}{2005}]{rines05} Rines K., Geller M.~J., Kurtz M.~J., Diaferio A., 2005, AJ, 130, 1482 

 \bibitem[\protect\citeauthoryear{Schaap \& van de Weygaert}{2000}]{schaap00} Schaap W.~E., van de Weygaert R., 
 2000, A\&A, 363, L29 
 \bibitem[\protect\citeauthoryear{Schawinski et al.}{2007}]{schawinski07} Schawinski K., et al., 2007, ApJS, 173, 512
 \bibitem[\protect\citeauthoryear{Schubert et al.}{2017}]{schubert17} Schubert E., Sander J., Ester M., Kriegel H.-P., Xu X., 2017, TODS, 
 Vol. 42, Issue 3, p 19
 \bibitem[\protect\citeauthoryear{Smith et al.}{2010}]{smith10} Smith R.~J., et al., 2010, MNRAS, 408, 1417 
 \bibitem[\protect\citeauthoryear{Smith, Lucey, \& Carter}{2012}]{smith12} Smith R.~J., Lucey J.~R., Carter D., 2012, MNRAS, 421, 2982
 \bibitem[\protect\citeauthoryear{Smith et al.}{2012}]{smith12b} Smith R.~J., Lucey J.~R., Price J., Hudson M.~J., Phillipps S., 2012, 
 MNRAS, 419, 3167  
 \bibitem[\protect\citeauthoryear{Sousbie}{2011}]{sousbie11} Sousbie T., 2011, MNRAS, 414, 350 
 \bibitem[\protect\citeauthoryear{Sousbie, Pichon, \& Kawahara}{2011b}]{sousbie11b} Sousbie T., Pichon C., Kawahara H., 2011,
  MNRAS, 414, 384 
 \bibitem[\protect\citeauthoryear{Sun et al.}{2010}]{sun10} Sun M., Donahue M., Roediger E., Nulsen P.~E.~J., Voit G.~M., Sarazin C.,
  Forman W., Jones C., 2010, ApJ, 708, 946 

 \bibitem[\protect\citeauthoryear{Tanimura et al.}{2017}]{tanimura17} Tanimura H., Hinshaw G., McCarthy I.~G., Van Waerbeke L.,
  Ma Y.-Z., Mead A., Hojjati A., Tr{\"o}ster T., 2017, arXiv, arXiv:1709.05024 
  \bibitem[\protect\citeauthoryear{Taylor}{2005}]{taylor05} Taylor M.~B., 2005, ASPC, 347, 29 
  \bibitem[\protect\citeauthoryear{Tejos et al.}{2016}]{tejos16} Tejos N., et al., 2016, MNRAS, 455, 2662 
 \bibitem[\protect\citeauthoryear{Tran et al.}{2009}]{tran09} Tran K.-V.~H., Saintonge A., Moustakas J., Bai L., Gonzalez A.~H., Holden B.~P.,
  Zaritsky D., Kautsch S.~J., 2009, ApJ, 705, 809 

 \bibitem[\protect\citeauthoryear{Van Waerbeke, Hinshaw, \& Murray}{2014}]{van14} Van Waerbeke L., Hinshaw G., Murray N., 2014, 
 PhRvD, 89, 023508 
 \bibitem[\protect\citeauthoryear{Verdugo et al.}{2012}]{verdugo12} Verdugo M., Lerchster M., B{\"o}hringer H., Hildebrandt H., Ziegler B.~L., 
 Erben T., Finoguenov A., Chon G., 2012, MNRAS, 421, 1949 

 \bibitem[\protect\citeauthoryear{Wang, Owen, \& Ledlow}{2004}]{wang04} Wang Q.~D., Owen F., Ledlow M., 2004, ApJ, 611, 821 
 \bibitem[\protect\citeauthoryear{Werner et al.}{2008}]{werner08} Werner N., Finoguenov A., Kaastra J.~S., Simionescu A.,
  Dietrich J.~P., Vink J., B{\"o}hringer H., 2008, A\&A, 482, L29 
 \bibitem[\protect\citeauthoryear{Wright et al.}{2010}]{wright10} Wright E.~L., et al., 2010, AJ, 140, 1868-1881 
 \bibitem[\protect\citeauthoryear{Wyder et al.}{2007}]{wyder07} Wyder T.~K., et al., 2007, ApJS, 173, 293 

 \bibitem[\protect\citeauthoryear{Zabludoff \& Mulchaey}{1998}]{zabludoff98} Zabludoff A.~I., Mulchaey J.~S., 1998, ApJ, 496, 39 
 \bibitem[\protect\citeauthoryear{Zhang et al.}{2013}]{zhang13} Zhang Y., Dietrich J.~P., McKay T.~A., Sheldon E.~S., Nguyen A.~T.~Q.,
  2013, ApJ, 773, 115 
 
\end{thebibliography}
\end{document}